\documentclass[twocolumn,prl,amsmath,superscriptaddress,nofootinbib,amssymb]{revtex4}
\usepackage{graphics}
\usepackage{amssymb}
\usepackage{graphicx}
\usepackage{times}
\usepackage{epsfig}          %% um eps-Dateien einzubinden (\epsfig{file=...})

\begin{document}

\title{Direct Observation of Interband Spin-Orbit Coupling in a Two-Dimensional Electron System}

\author{Hendrik~Bentmann}\affiliation{Experimentelle Physik VII and R\"ontgen Research Center for Complex Materials (RCCM), Universit\"at W\"urzburg Am Hubland, D-97074 W\"urzburg, Germany}\affiliation{Karlsruhe Institute of Technology KIT, Gemeinschaftslabor f\"ur Nanoanalytik, D-76021 Karlsruhe, Germany}
\author{Samir~Abdelouahed}\affiliation{Texas A\&M University at Qatar, P.O. Box 23874, Doha, Qatar}
\author{Mattia~Mulazzi}\affiliation{Experimentelle Physik VII and R\"ontgen Research Center for Complex Materials (RCCM), Universit\"at W\"urzburg Am Hubland, D-97074 W\"urzburg, Germany}\affiliation{Karlsruhe Institute of Technology KIT, Gemeinschaftslabor f\"ur Nanoanalytik, D-76021 Karlsruhe, Germany}
\author{J\"urgen~Henk}\affiliation{Institut f\"ur Physik - Theoretische Physik, Martin-Luther-Universit\"at Halle-Wittenberg, D-06099 Halle (Saale), Germany}
\author{Friedrich Reinert}\affiliation{Experimentelle Physik VII and R\"ontgen Research Center for Complex Materials (RCCM), Universit\"at W\"urzburg Am Hubland, D-97074 W\"urzburg, Germany}
\affiliation{Karlsruhe Institute of Technology KIT, Gemeinschaftslabor f\"ur Nanoanalytik, D-76021 Karlsruhe, Germany}

\date{\today}
\begin{abstract}
We report the direct observation of interband spin-orbit (SO) coupling in a two-dimensional (2D) surface electron system, in addition to the anticipated Rashba spin splitting. Using angle-resolved photoemission experiments and first-principles calculations on Bi/Ag/Au heterostructures we show that the effect strongly modifies the dispersion as well as the orbital and spin character of the 2D electronic states, thus giving rise to considerable deviations from the Rashba model. The strength of the interband SO coupling is tuned by the thickness of the thin film structures. 
\end{abstract}
\maketitle
The spin-orbit interaction plays a fundamental role in the rapidly developing field of spintronics as it allows for the electrostatic manipulation of the spin degrees of freedom in the conduction channels of nanoscale heterostructures \cite{Zutic:04.4,koo:09}. Such applications are often based on the Rashba effect in a two-dimensional electron gas (2DEG) 
that prescribes a lifting of the spin degeneracy in the presence of structural inversion asymmetry and strong spin-orbit (SO) coupling \cite{Rashba:84}. Other contributions to the spin splitting arise from the Dresselhaus effect for constituent bulk crystal structures without a center of inversion \cite{Dresselhaus:55.10}. Moreover, new effects due to the SO interaction have been discovered recently that give rise to topologically protected, spin-polarized states on the surfaces of a number of heavy-element semiconductors, referred to as topological insulators \cite{Fu:07.7,Hasan:09.2}.

In the phenomenological treatment of the mentioned effects the spin-polarized electronic states are usually assumed as pure spin states. In a real system, on the other hand, the SO interaction couples spin and orbital angular momentum which will result in a mixing of orthogonal spinors in the single-particle eigenstates \cite{henk:96.1}. Recent \textit{ab-initio} calculations suggest that this can result in considerable reductions of the spin polarization of spin-split two-dimensional electronic states in the presence of strong SO interaction \cite{Louie:10.12}. Even more profound effects of spin-mixing are known from the three-dimensional (3D), exchange-split band structures of ferromagnets where the SO coupling allows for hybridizations between spin-up and spin-down bands. This interband SO coupling lies at the origin of several magnetic phenomena, e.g. magneto crystalline anisotropy \cite{Bruno:89.1} or ultrafast demagnetization \cite{Pickel:08.8}. Note that spin-mixing is expected to gain increasingly in importance for materials with high atomic number and thus enhanced SO interaction. It is therefore of fundamental importance to explore whether interband SO coupling effects due to spin-mixing are present in heavy-element 2DEGs and how they modify the spin-split electronic structure. Indeed, recent theoretical reports predict interband SO coupling phenomena in 2DEGs formed in zinc blende quantum wells \cite{Loss:07.8,Loss:08.10} and on high-\textit{Z} metal surfaces \cite{Henk:09.06} as a result of higher-order perturbation theory corrections in the SO interaction. Yet, to best of our knowledge, these effects have not been addressed in experiment so far.

\begin{figure}[t]
\includegraphics[width=3.2in]{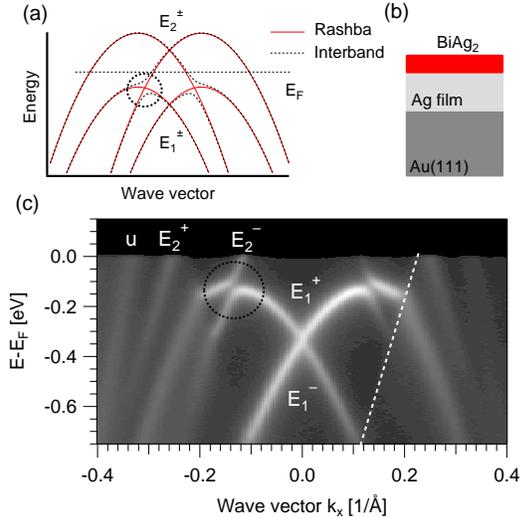}
\caption{\label{fig1}(Color online) (a) Band structure of a two-band system with Rashba interaction (schematic). Two different cases with (dashed line) and without (full line) an additional interband spin-orbit coupling are shown. (b) Sketch of the studied multilayer sample: The Bi$\mbox{Ag}_{2}$ layer is grown on a Ag film of variable thickness supported by the Au(111) substrate. (c) Experimental band structure for Bi$\mbox{Ag}_{2}$ on 4~ML~Ag/Au(111) along the $\bar\Gamma$$\bar K$ direction of the surface Brillouin zone. The dashed line indicates the edge of the projected bulk band gap of Au(111).}
\end{figure}

In this Letter we report the direct observation of interband SO coupling in a 2DEG with large Rashba splitting that is formed in a Bi$\mbox{Ag}_{2}$ surface alloy grown on Ag quantum films supported by a Au(111) substrate [Fig.~\ref{fig1}(b)]. Using angle-resolved photoelectron spectroscopy (ARPES) with high energy-resolution we find avoided crossings in the spin-split electronic structure that provide evidence for the hybridization of states with opposite spin due to SO coupling. These findings are in line with relativistic first-principles computations and model calculations. Further, we demonstrate that the strength of the interband SO effect varies upon changing the thickness of the Ag film whereas the Rashba coupling remains unmodified. The new interband coupling hence emerges to be tunable, independently from other SO effects, by nanostructural design capabilities as shown here for the layered Ag/Au heterostructure. Our investigations prove that the SO interaction can considerably influence the electronic states in 2D systems, in addition to established effects such as Rashba splitting. We hence expect the present findings to be highly relevant for spintronic applications based on strongly SO coupled compounds including the recently reported heavy-element semiconductors with large Rashba splitting \cite{Kimura:11.6,hofmann:11.3} and topological insulators \cite{xia_observation_2009}.       

The Bi$\mbox{Ag}_{2}$ system and related isostructural alloys have been shown previously to feature an unusually large Rashba effect in their electronic structure resulting from the strong SO interaction of the Bi atoms \cite{pacile,ast,Moreschini:09,bentmann:09}. The band structure of the surface alloys consists of two parabolic, Rashba-split states  $E^{\pm}_{1,2}$ with negative effective mass [Fig.~\ref{fig1}(a)], whose spin-polarization has been verified by spin-resolved photoemission experiments \cite{meier,He:10.4,Bentmann:11.9}.

The ARPES data were collected by a SCIENTA R4000 electron spectrometer employing a monochromatized He discharge lamp operating at an excitation energy of 21.22~eV (He~$\mbox{I}_{\alpha}$). The energy and angular resolution of the setup are $\Delta$\textit{E}\,=\,3\,meV and $\Delta\theta$\,=\,0.3$^\circ$, respectively. We performed all measurements at base pressures lower than $2\cdot10^{-10}$ mbar and at temperatures of 20~K. The preparation of the Au(111) substrate by cycles of Ar sputtering and annealing resulted in a clean and well-ordered surface as verified by the spectral linewidth of the Au(111) surface state \cite{reinert}. Ag was evaporated on the cooled substrate at $\sim$150~K. Subsequent mild annealing resulted in homogeneous films as confirmed by the characteristic shift in binding energy of the surface state for low Ag coverages \cite{Popovic:05.7}. The Bi$\mbox{Ag}_{2}$ surface alloy was obtained after evaporation of 1/3~ML Bi on the substrate film at elevated temperatures of 400~K. We observed no change in the binding energy of the Shockley state for Ag/Au(111) at this temperature providing evidence for no or neglegible Ag/Au intermixing at the interface \cite{Cercellier:06.5}. Low energy electron diffraction measurements confirmed the ($\sqrt{3}\mbox{x}\sqrt{3}$) reconstruction of the surface alloy.

First-principles calculations were carried out using the relativistic full-potential linearized augmented plane wave (FLAPW) method as implemented in the Fleur code \cite{FLEUR}. Exchange and correlation were treated within the generalized gradient approximazion (GGA) \cite{perdew_96.10}.
We used a plane-wave cutoff of 8.6~\AA$^{-1}$ while the charge density and potential cutoffs were 21.9~\AA$^{-1}$. A ten-layer slab (Bi$\mbox{Ag}_{2}$/4ML Ag/5 ML Au) was used to simulate the surface electronic structure. The opposite slab surface was saturated by hydrogen atoms to suppress the formation of the Au(111) surface state. The vertical relaxation of the Bi atoms was 0.95~\AA.  

Fig.~\ref{fig1}(c) shows the experimental electronic structure of a Bi$\mbox{Ag}_{2}$ surface alloy grown on a 4~ML Ag film on Au(111). We identify two pairs of states $E^{\pm}_{1,2}$, both showing Rashba spin-splitting around the $\bar\Gamma$ point, and a backfolded Au bulk band $u$. The measured surface band structure thus complies with the scheme in Fig.~\ref{fig1}(a) with a position of the Fermi energy as indicated. Deviations from the pure Rashba scenario [full line in Fig.~\ref{fig1}(a)] are observed in the experimental data at the expected crossing points between the two branches $E^{+}_{1}$ and $E^{-}_{2}$. Near these points the two states hybridize which leads to a gap opening and pronounced, kink-like changes in the dispersion [highlighted by the circle in (c)]. Note that these observations cannot be explained within the Rashba model which assigns pure opposite spin states to these two bands. A hybridization between them should be prohibited in this case. The findings are therefore distinctively different from the previously found hybridization phenomena between Rashba-split states and spin-degenerate quantum well states \cite{He08,Grioni:08,He:10.4}. Our experimental results indicate an additional interband SO effect that couples orbitals with opposite spin and thereby induces the hybridization.

\begin{figure}[b]
\includegraphics[width=3.4in]{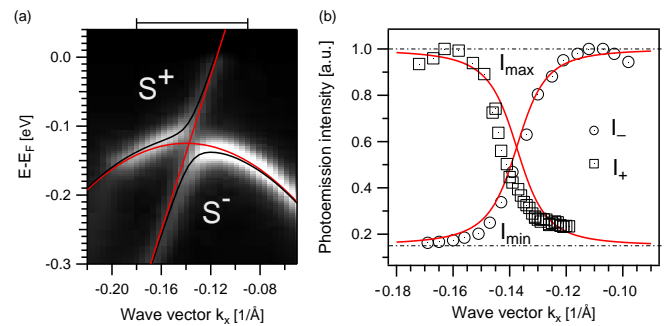}
\caption{\label{fig2}(Color online) Comparison of the experimental electronic structure close to the SO-induced hybridization gap with the solution of a model Hamiltonian (see text). A closeup of the hybridization gap taken from the dataset in Fig.~\ref{fig1}(c) is shown in (a). The ARPES spectrum is overlaid by the modelled dispersions $S_{\pm}$ (dark lines) for an interaction strength $\Delta$~=~31~meV. The scale bar above the graph indicates the $k$-interval of panel (b). In (b) the datapoints mark the measured $k$-dependent intensity evolution  $I_{\pm}$ of the two branches $S_{\pm}$. The experimental results are compared with the calculated intensities (full lines) based on the model Hamiltonian.}  
\end{figure}

To gain a first understanding of the observed effect we adopt an effective 2D model Hamiltonian taking into account Rashba and interband contributions of the SO coupling but neglecting the orbital part of the wave functions for the moment, similar as it is done for the Rashba model in single band systems. The interband SO coupling term introduces a finite hybridization $\Delta$ = $\langle E^{+}_{1} | H_{so} | E^{-}_{2} \rangle$ between the two purely Rashba-split states. For the modified eigenvalues $S_{\pm}$ we then have $S_{\pm}=\frac{1}{2}(E^{+}_{1}+E^{-}_{2})\pm \sqrt{\frac{1}{4}(E^{+}_{1}-E^{-}_{2})^2+\Delta^2}$, where we omit the $k$-dependence for clarity. Close to the  crossing point of $E^{+}_{1}$ and $E^{-}_{2}$ the SO coupling mixes the states $|E^{+}_{1}\rangle$ and $|E^{-}_{2}\rangle$ resulting in new eigenstates 
$|S_{+} \rangle= a_k |E^{+}_{1}\rangle + \sqrt{1-a_k^2} | E^{-}_{2} \rangle$ and $| S_{-} \rangle = \sqrt{1-a_k^2}  |E^{+}_{1}\rangle - a_k| E^{-}_{2} \rangle$. The coefficient $a_k$ can be expressed by $|a_k|^2=(1+\frac{(S_{+}-E^{+}_{1})^2}{\Delta ^2})^{-1}$.

Our experimental data allow us not only to determine the modified dispersions $S_{\pm}$ but also to trace the $k$-dependence of the coefficient $|a_k|^2$ [Fig.~\ref{fig2}]. The latter is possible due to considerably differing photoionization cross sections for the two states $|E^{+}_{1}\rangle$ (high cross section) and $|E^{-}_{2}\rangle$ (low cross section). As a result, the branches $S_{\pm}$ show drastic intensity changes close to the hybridization gap where their state character is strongly modified. We identify the $k$-dependent intensity evolution $I_{\pm}$ of the branches $S_{\pm}$ with the relative contributions of $|E^{+}_{1}\rangle$ and $|E^{-}_{2}\rangle$ to $|S_{\pm}\rangle$. These contributions are directly given by $a_k$ and we have the simple correspondence $I_{+}=I_{min}+(I_{max}-I_{min})|a_k|^2$ and accordingly for $I_{-}$. Within this approximation we neglect photoemission matrix element effects. Note that $a_k$ and $I_{\pm}$ depend on the hybridization strength $\Delta$.

\begin{figure}[t]
\includegraphics[angle=-90,width=3.2in]{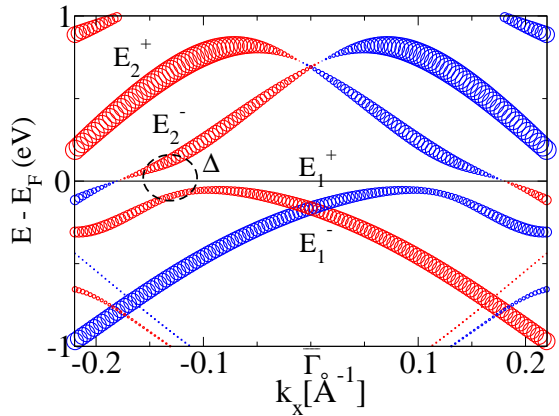}
\caption{\label{fig3}(Color online) Spin-resolved surface electronic struture of Bi$\mbox{Ag}_{2}$ on 4~ML~Ag/Au(111) as obtained by a first-principles calculation. The size of the markers scales with the degree of spin-polarization of the electronic states and the color refers to the spin-orientations up (red,light) and down (blue,dark) with the spin-quantization axis along the $y$-direction. The hybridization gap $\Delta$ between the branches $E^{+}_{1}$ and $E^{-}_{2}$ is indicated.}
\end{figure}

A comparison of the experimental data with the results of the model Hamiltonian is shown in Fig.~\ref{fig2}. In (a) we display the measured band structure and the model dispersions close to the hybridization gap. The bands $E^{+}_{1}$ and $E^{-}_{2}$ [red (light) lines in Fig.~\ref{fig2}(a)] are determined from the experimental data in ($E,k$) regions sufficiently far away from the hybridization gap. For $E^{-}_{2}$ we use a linear dispersion which is an adequate approximation for the small ($E,k$) window which is of interest here. A close match between the experimental bands and $S_{\pm}$ is obtained for $\Delta$~=~31~meV. Fig.~\ref{fig2}(b) shows the photoemission intensities $I_{\pm}$ obtained from energy and momentum distribution curves as a function of the wave vector $k_x$. The data for each branch were normalized to the respective maximal value of the individual data set. For increasing wave vector $k_x$ the intensity $I_{+}$ is enhanced whereas $I_{-}$ is reduced reflecting the change in state character of both branches near the hybridization gap. We confirmed that this behavior is not affected by the finite \textit{k}-resolution of the experiment ($\sim$0.01~\AA$^{-1}$). The full lines in (b) correspond to the expected intensity change according to the model Hamiltonian using the same value for $\Delta$ and the same dispersion relations as in (a). Again, we find a good agreement with the experimental data. We hence conclude that the observed dispersion modification and change in state character can be consistently described by a Rashba-type free electron model including an additional interband SO coupling term. Comparing the energy scales of the two SO effects in this system we find the interband contribution ($\sim$30~meV) to be about one order of magnitude weaker than the Rashba contribution ($\sim$300~meV).

\begin{figure}[b]
\includegraphics{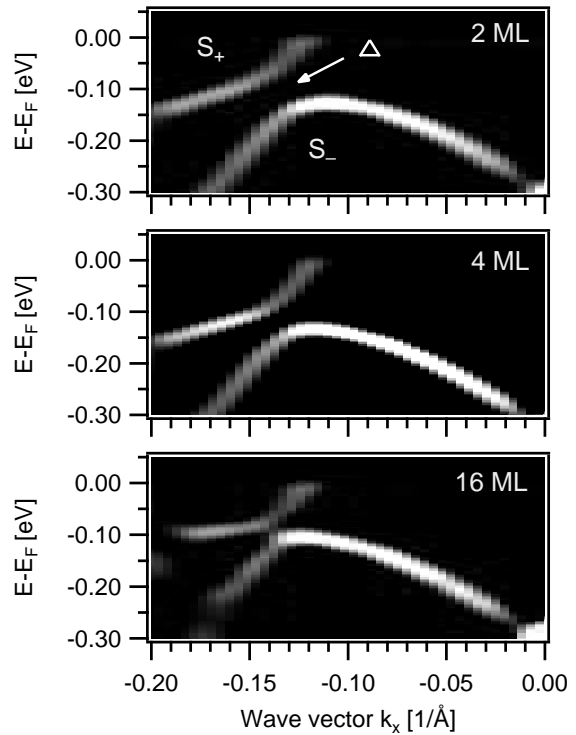}
\caption{\label{fig4}Modification of the SO-induced hybridization gap $\Delta$ arising from changes in the thickness of the supporting Ag film. The three panels show second-derivative spectra of the experimental data for Ag layers of 2~ML, 4~ML and 16~ML.}
\end{figure}

Qualitative insight into the coupled orbital and spin parts of the electronic states $|E^{+}_{1}\rangle$ and $|E^{-}_{2}\rangle$ which goes beyond the discussed free electron model can be attained by inspecting the spatial symmetry and the contributing atomic orbitals in the system. Group-theoretical considerations show that both states belong to the same representation and that their wave function along $k_x$ can be written as $|\psi\rangle=|sp_{z}\uparrow\rangle+|p_{x}\uparrow\rangle+|p_{y}\downarrow\rangle$, where the spinors are quantized with respect to the $k_y$ axis \cite{Henk:09.06}. It is important to note that the SO interaction induces a mixing of opposite spinors in the wave function and therefore facilitates a hybridization between the two bands which is indeed observed in our measurements. The hybridization mechanism is therefore comparable to the effect of SO coupling on the 3D spin-split band structures in ferromagnetic systems \cite{Henk:98.6,Pickel:08.8}. 

To further substantiate the experimental results on a quantitative theoretical level we examine the spin-polarized electronic structure by a first-principles calculation [Fig.~\ref{fig3}]. In accordance with experiment our calculation finds two Rashba-split states and a hybridization gap between the two branches $E^{+}_{1}$ and $E^{-}_{2}$. Similar to the measured results the strength of interband coupling is smaller than that of the Rashba coupling. Note that the spin-polarization of the branch $E^{-}_{2}$ shows a sign change near the hybridization kink clearly reflecting the mixed and $k$-dependent spin character of the corresponding wave function. We further find the degree of spin-polarization to be reduced from 100~\%, as predicted by the Rashba model, to maximal values of $\sim$66~\%. 

It is interesting to explore the possibility to modify the electronic structure of the Bi$\mbox{Ag}_{2}$ alloy by choosing a different thickness for the supporting Ag quantum well film. We have recently shown that the spatial localization of quantum well states in the system Ag/Au(111) changes considerably as a function of layer thickness, especially in the low coverage regime up to $\sim$20~ML \cite{Forster:11.8}. As a result the charge density close to the surface, where the Bi$\mbox{Ag}_{2}$ alloy is located, varies strongly depending on the precise Ag film thickness. Hence, one may expect concomitant influences of these variations on the electronic states in the surface alloy. Indeed, we find such changes as is inferred from Fig.~\ref{fig4} which shows the electronic structure of Bi$\mbox{Ag}_{2}$ close to the hybridization gap for three different Ag layers. Whereas the general features in the electronic structure, such as binding energy and Rashba splitting, are very similar for all three films we find considerable changes in the size of the hybridization gap $\Delta$ and therefore in the interband coupling strength. The largest gap is found for 2~ML ($\Delta$~=~42~meV) whereas at higher thicknesses it is reduced ($\Delta$~=~14~meV for the case of 16~ML). The strength of the interband SO coupling is therefore determined by details of the geometric and electronic substrate properties. This provides possibilities to tailor this new SO effect by a controlled variation of the corresponding parameters in custom designed nanostructures, as exemplified here by the electronically tunable quantum well system Ag/Au.  

We have established the presence of interband SO coupling in 2DEGs with Rashba spin splitting by photoemission spectroscopy experiments on multilayered Bi/Ag/Au heterostructures. The coupling induces hybridization between bands of different spin polarizations and thereby causes considerable deviations from the Rashba model in the dispersion of the 2D electronic states as well as in their spin and orbital momentum character. Such effects will certainly influence the transport properties in 2DEGs, especially in compounds with very strong SO interactions, and may therefore be exploited to improve the performance of new spintronic functionalities \cite{Loss:03.2,Loss:02.10}.

HB acknowledges helpful discussions with Frank Forster and Kazuyuki Sakamoto. This work was supported by the Bundesministerium f\"ur Bildung und Forschung (Grants No. 05K10WW1/2 and No. 05KS1WMB/1) and the Deutsche Forschunsgsgemeinschaft within the Forschergruppe 1162 and the Sonderforschungsbereich 762.
\bibliographystyle{apsrev}

\end{document}